\documentclass[runningheads]{llncs}
\usepackage[colorinlistoftodos]{todonotes}
\usepackage[T1]{fontenc}
\usepackage{graphicx}
\usepackage{subcaption}
\usepackage{hyperref}
\usepackage{color}
\renewcommand\UrlFont{\color{blue}\rmfamily}

\usepackage{amsmath, amssymb}

\begin{document}
\title{VADA: a Data-Driven Simulator for Nanopore Sequencing}
%
%

\author{Jonas Niederle\inst{1}\orcidID{0009-0009-9854-0489} \and
Simon Koop\inst{1}\orcidID{0009-0003-2673-6104} \and
Marc Pagès-Gallego \inst{2}\inst{3}\orcidID{0000-0001-8888-5699} \and
Vlado Menkovski \inst{1}\orcidID{0000-0001-5262-0605}}
\authorrunning{J. Niederle et al.}
%
\institute{Eindhoven University of Technology, Eindhoven, The Netherlands \and
Oncode Institute, Utrecht, The Netherlands \and Center for Molecular Medicine, UMC Utrecht, Utrecht, The Netherlands
}
\maketitle              
\begin{abstract}

Nanopore sequencing offers the ability for real-time analysis of long DNA sequences at a low cost, enabling new applications such as early detection of cancer. Due to the complex nature of nanopore measurements and the high cost of obtaining ground truth datasets, there is a need for nanopore simulators. Existing simulators rely on handcrafted rules and parameters and do not learn an internal representation that would allow for analysing underlying biological factors of interest.
Instead, we propose VADA, a purely data-driven method for simulating nanopores based on an autoregressive latent variable model. We embed subsequences of DNA and introduce a conditional prior to address the challenge of a collapsing conditioning. We experiment with an auxiliary regressor on the latent variable to encourage our model to learn an informative latent representation. We empirically demonstrate that our model achieves competitive simulation performance on experimental nanopore data. Moreover, we show our model learns an informative latent representation that is predictive of the DNA labels. We hypothesize that other biological factors of interest, beyond the DNA labels, can potentially be extracted from such a learned latent representation.


\end{abstract}

\section{Introduction}
DNA contains the genetic instructions needed for all living organisms to grow, reproduce, and function. Nanopore sequencing is an emerging DNA sequencing technique, which allows real-time analysis of long sequences of DNA, has low costs, is portable, and requires little preparation time, in steep contrast to traditional DNA sequencing approaches, which are costly, can only process short sequences of DNA and require much more preparation and processing time.
These advantages make nanopore sequencing suitable to, for example, be used for early detection and treatment of cancer \cite{Norris2016NanoporeCancer,Lin2021NanoporeSequencing}. Furthermore, during a pandemic, nanopore sequencing can be used for rapid detection of virus mutations \cite{Lin2021NanoporeSequencing}. 


Nanopore sequencing works by applying current to a tiny hole, a nanopore, passing a sequence of DNA through it, and measuring the resulting change in current. DNA \textit{bases}, A, C, G, and T, make up the individual elements of a DNA sequence, and, a \textit{k-mer} refers to a subsequence of length \textit{k}. Each DNA base affects the current signal differently, thus capturing the change in electrical current allows the identification of the DNA sequence. 

Determining the DNA sequence from the current measurements is challenging because of several reasons. Firstly, multiple bases are in the nanopore simultaneously. Therefore, the change in observed electrical current depends on multiple bases, i.e. a k-mer \cite{Pages-Gallego2023ComprehensiveBasecalling}. Secondly, the DNA sequence moves through the nanopore at a non-constant speed \cite{Pages-Gallego2023ComprehensiveBasecalling}. This causes variability in the number of nanopore measurements that corresponds to one k-mer. Therefore, multiple \textit{timewarped} versions of a current sequence can occur. 

Lastly, besides the sequence of bases, other exogenous variables influence the observed current. For example, additional chemical modifications, such as methylation, can occur on top of the four canonical bases, influencing the biological function of DNA and the resulting current. 

\begin{figure}[t]
\centering
\includegraphics[width=\textwidth]{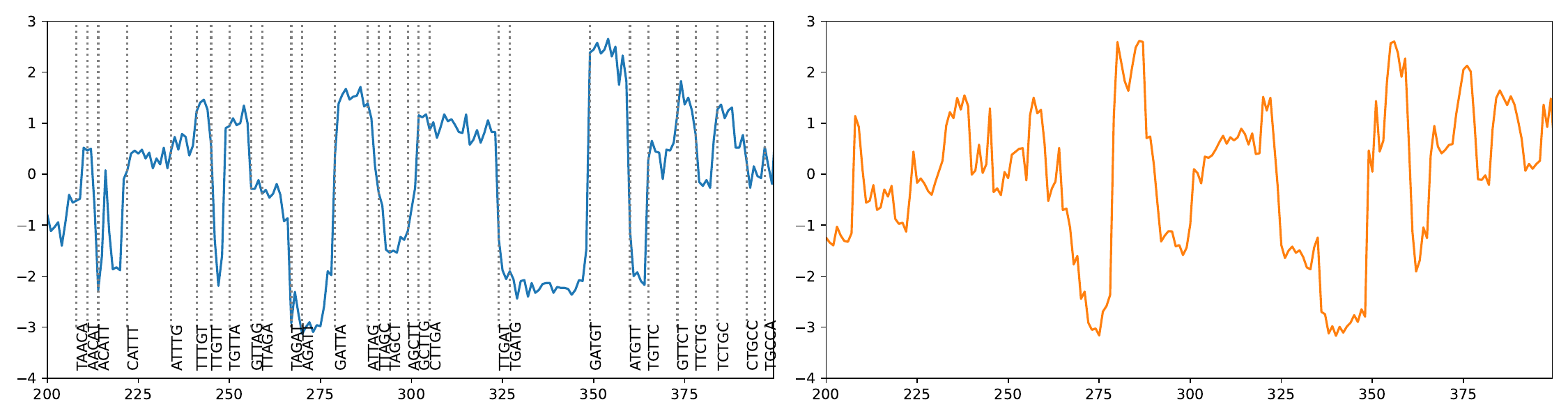}
\caption{An example of experimental nanopore signal (left) with aligned k-mers and generated signal for the same k-mer alignment (right).} \label{fig:real_generated}
\end{figure}

The left plot in Figure \ref{fig:real_generated} shows an example of Nanopore signal with an aligned sequence of k-mers. Here one can clearly see the variability in nanopore current and speed of DNA moving through the nanopore.

As a result of this complexity, Machine Learning-based predictive models are used to determine the sequence of bases from the raw current measurements. This process is referred to as \textit{basecalling} and is an essential step for most downstream applications \cite{Wang2021NanoporeApplications}.

Besides basecalling, new methods are rapidly being developed to analyse nanopore data for downstream tasks \cite{Deamer2016ThreeSequencing,Li2018DeepSimulator:Sequencing}. Such new methods require labeled data for benchmarking and potentially also for training. Although evaluation on empirical data is important to guarantee the quality of any method, such empirical data is costly and ground truth data is hard to come by: obtaining ground truth labels involves sequencing the same DNA using an orthogonal sequencing method. Simulating nanopore signals and the resulting base calls allows for cheap supplementation of empirical data for more extensive benchmarking.

To address this need, multiple simulators have been developed \cite{Li2018DeepSimulator:Sequencing,Li2020DeepSimulator1.5:Sequencing,Chen2020SimulationBiGRU,Rohrandt2019NanoporeSequencing}. 
These approaches are typically implemented in two separate steps. First, a \textit{deterministic} estimation of the expected nanopore current is produced for each k-mer. Then, Gaussian noise is added to the expected current to effectively produce a sample from a probability distribution that governs the simulation process. Importantly, the standard deviation of the noise term is often \textit{user-defined} and \textit{constant} for all k-mer \cite{Chen2020SimulationBiGRU,Li2020DeepSimulator1.5:Sequencing} or estimated by a Gaussian distribution \cite{Rohrandt2019NanoporeSequencing}, and always \textit{independent} of the number of consecutive k-mer measurements and any other context. 

These assumptions about the variance in the observations do not match what we observe in experimental data. The distribution of nanopore current measurements varies greatly per k-mer, 
it is not normally distributed, and it depends on the number of consecutive measurements of the k-mer. 

This is further illustrated by the fact that errors of discriminative basecalling models occur more frequently for some k-mers than for others \cite{Pages-Gallego2023ComprehensiveBasecalling,Delahayeid2021SequencingBiases}. So, existing approaches are fundamentally incapable of modeling the variability that is observed in the data. 

Moreover, capturing all the sources of variation in the data is not only essential for effectively simulating nanopore sequencing, but a model that does so, can also be of use to biologists for analysing these sources of variation. For example, nanopore sequencing has been used for the detection of DNA methylation \cite{Wang2021NanoporeApplications}. Existing approaches do not offer the ability for further analysis of underlying and potentially unknown patterns in the DNA.

To address these limitations, we develop a \textit{data-driven nanopore simulator} based on a deep generative model. The main goal of our model is to capture the variability in nanopore current measurements that correspond to the DNA bases. Therefore, we aim to model the \textit{distribution over nanopore current sequences}, conditioned on a given DNA sequence. As we aim to efficiently simulate the nanopores, our model needs to allow for efficient sampling from this distribution. Importantly, in contrast to current approaches, a data-driven simulator must learn to model the stochastic process of nanopore sequencing exclusively from data, and thus cannot rely on the estimation of deterministic values or make assumptions on the shape of the distribution of nanopore currents. 

Accordingly, we propose a latent variable model similar to a Variational Autoencoder \cite{Kingma2013Auto-EncodingBayes} and DIVA \cite{Ilse2020DIVA:Autoencoders}. By introducing a latent variable, we can model high dimensional, complex distributions of arbitrary shape, while enabling us to efficiently sample multiple nanopore observations by sampling from the latent space. 

To condition our model in accordance with the physical properties of nanopore sequencing, we represent the DNA sequence by embedding k-mers of DNA, as done in other machine learning tasks in this domain \cite{Trabelsi2019ComprehensiveSpecificities}. Initial empirical results showed that a straightforward approach to conditioning the latent variable model results in a \textit{conditioning collapse}, where the model ignores the conditioning and produces bad samples unrelated to the DNA sequence.
%
To overcome this problem, and effectively condition our model on a sequence of DNA, we introduce a conditional prior distribution on the latent space. 

In this work, we propose a data-driven nanopore simulator based on a deep generative model. We summarize our contributions as follows:

\begin{itemize}
    \item We propose the Variational Autoregressive DNA-conditioned Autoencoder (VADA), an autoregressive probabilistic model for data-driven simulation of nanopore sequencing. 
    \item We show VADA can effectively model DNA-conditioned probability distributions over nanopore current sequences to produce varying current observations, and does so by exclusively learning from data.     
    \item We evaluate VADA on publicly available experimental nanopore data, which was obtained by sequencing human DNA. We show our results are competitive to a non-data driven approach. 
    \item We show VADA learns a meaningful representation of nanopore current sequences that can be used for analysis, by training a classifier on samples from the approximate posterior on the latent space, and demonstrating that we can accurately determine the DNA bases that produced the nanopore current sequence.  
\end{itemize}

\section{Methods}

The simulation of nanopore sequencing can be described in terms a sequence of nanopore current measurements $x^0, \dots, x^{T}$ and an aligned sequence of DNA k-mers $y^0, \dots, y^{T}$ as sampling from a distribution $p(x^0,\ldots, x^T\mid y^0,\ldots, y^T)$ of current measurements conditioned on the aligned sequence of k-mers. Here, $y^t$ denotes the k-mer that was (in the center of) the nanopore at time $t$ and corresponds to nanopore measurement $x^t$. Our goal is to learn this distribution from data.
The DNA sequence is represented as a sequence of 5-mers, as we know that approximately 5 bases are in the nanopore simultaneously and because pragmatically, there is a center DNA base in the k-mer. As an example, a single sample at time $t$ might be described by $x^t=-0.4$ and $y^t=CATCG$. 

\subsection{VADA: Variational Autoregressive DNA-conditioned Autoencoder}
We are interested in modeling a distribution over nanopore observations for a given DNA sequence. We approach this task by modeling windows of nanopore current sequences of length $\delta$. We know that nanopore measurements are predominantly influenced by the k-mer currently in the pore. Additionally, because nanopore sequencing is a continuous process and DNA bases at the edge of the k-mer might be only partly in the pore, the previous window of nanopore measurements $x^{t-\delta:t}$ will affect the current window of measurements $x^{t:t+\delta}$. Therefore, we model the distribution $p(x^{0:T} \mid y^{0:T})$ autoregressively as a product of distributions over windows:
\begin{equation}    \label{eq:modelling_task}
    p_{\theta}(x^{t:t+\delta}|y^{t:t+\delta}, x^{t-\delta:t})
\end{equation}

Where $\theta$ are the parameters of the model. For the initial window, we learn a separate distribution as $p_\theta(x^{0:\delta}\mid y^{0:\delta})$.



\subsubsection{Latent Variable Model}
Our goal is to model a high-dimensional, complex distribution over windows of nanopore current sequences, without making any assumptions about the shape of the distribution. As we know the same sequence of DNA can result in a diverse set of nanopore current observations, and the model needs to capture this variability. At the same time, we want to efficiently sample multiple observations from this distribution.

Therefore, we model the problem using a latent variable model \cite{Kingma2013Auto-EncodingBayes}. This type of model can represent a complex distribution in a high dimensional space while allowing for efficient sampling of nanopore observations, by sampling from the latent space. Specifically we model Equation \ref{eq:modelling_task} using latent variable $z$ as $\int p_\theta(x^{t:t+\delta}|z, x^{t-\delta:t})p_\theta(z|y^{t:t+\delta})dz$, where $p_\theta(x^{t:t+\delta}|z, x^{t-\delta:t})$ and $p_\theta(z|y^{t:t+\delta})$ are parameterized by neural networks.

\subsubsection{DNA representation} Multiple DNA bases influence each nanopore current measurement.  Consequently, as mentioned earlier, we represent $y_t$ as a k-mer containing $k=5$ DNA bases. The embedding layer $f_\theta$ processes each $y_t$ in the window independently, producing embeddings $e^{t:t+\delta}=f_\theta(y^{t:t+\delta})$.

\begin{figure}[t]
\includegraphics[width=\textwidth]{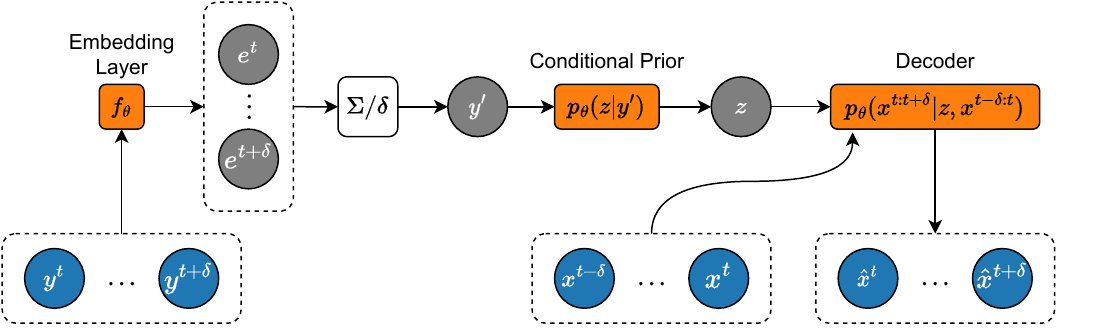}
\caption{VADA sampling overview.} \label{fig:vada_sampling}
\end{figure}

\subsubsection{Conditioning}
To simulate nanopore current sequences specific to a DNA sequence, we need to condition the model. But, a straight-forward approach where the model is conditioned via the approximate posterior, i.e. the encoder, and the decoder \cite{Doersch2016TutorialAutoencoders} results in a \textit{conditioning collapse}. Meaning, the model largely ignores the DNA sequence and only learns to reconstruct from a latent sample of the approximate posterior,  but cannot produce realistic samples for a \textit{given} DNA sequence by sampling from the prior distribution.

Additionally, we know the same sequence of DNA bases can result in strongly varying nanopore currents. Therefore, to enable our model to simulate this behavior and to overcome the problem of conditioning collapse, we condition our model via the prior distribution \cite{Ilse2020DIVA:Autoencoders}. Specifically, the embedded k-mers are summed and averaged over the window producing $y'=\sum_{k=t}^{t+\delta-1}e^k / \delta$. Subsequently, $y'$ is used to condition the prior distribution $p_\theta(z|y')$. We learn the prior distribution, which is modeled as $p_\theta(z|y')=\mathcal{N}(\mu^\text{prior}(y'),\sigma^\text{prior}(y'))$, where $\mu^\text{prior}$ and $\sigma^\text{prior}$ are neural networks.

Experimental nanopore sequencing data contains an alignment of each nanopore current measurement to a specific k-mer. However, this alignment is created by using an optimization algorithm \cite{Pages-Gallego2023ComprehensiveBasecalling}. Moreover, perfectly aligning nanopore currents measured at fixed time intervals is not possible due to the continuous nature of nanopore sequencing. Thus, despite the alignment, there is variability in the correspondence of k-mer with the nanopore current sequence. Accordingly, by aggregating the conditioning we enable our model to simulate nanopore current observations with variability in time alignment of the k-mer sequence.

After conditioning, the latent sample $z\sim p_\theta(z|y')$ is processed by the decoder together with the previous window of observations $x^{t-\delta:t}$, to produce a distribution $p_\theta(x^{t:t+\delta}|z, x^{t-\delta:t})$ over the next window of nanopore observations. The entire sampling process is visualized in Figure \ref{fig:vada_sampling}.

\subsubsection{Informative latent space}
Our model should not only be able to simulate nanopore current observations, but should also allow for analysis of the underlying sources of variation.

Our model inherently learns a latent representation of the nanopore currents via the latent variable $z$ and is further encouraged to encode information about the DNA sequence into $z$ through the use of the conditional prior \cite{Ilse2020DIVA:Autoencoders}. Nevertheless, there is no guarantee that $z$ captures the true underlying sources of variation, such as the DNA sequence that was in the pore. 

\begin{figure}[t]
\centering
\includegraphics[width=\textwidth]{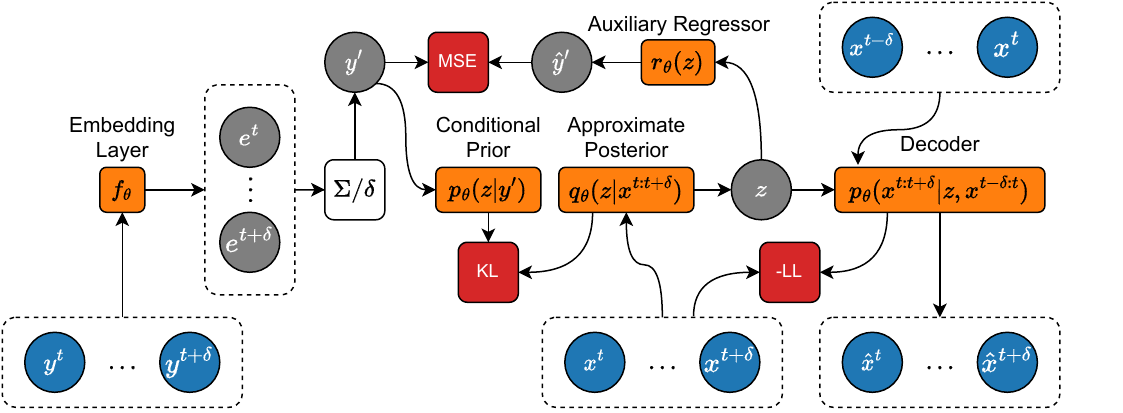}
\caption{VADA training overview, including the additional regressor used for experimentation in Section \ref{sec:analysis_of_latent_space}. \label{fig:vada_training_complete}} 
\end{figure}

To show that this is indeed the case, we experiment with a modified version of our model which is encouraged tolearn a representation that contains information about the DNA sequence through use of an auxiliary regressor during training, as done in other VAE's that aim to learn an informative latent space, such as DIVA \cite{Ilse2020DIVA:Autoencoders}. Specifically, during training, this regressor processes a latent sample $z$ to predict the aggregated embedding, $y'$, that was used to condition the prior, as $\hat{y}'=r_\theta(z)$. This auxiliary regressor is included in the visualization of the training procedure (Fig. \ref{fig:vada_training_complete}). 

In Section \ref{sec:analysis_of_latent_space}, we show that the model without this additional regressor performs as well as the model with the added regression objective, showing that the conditional prior indeed learns informative latent representations of the various k-mers. 



\subsubsection{Training} 
Following the VAE framework \cite{Kingma2013Auto-EncodingBayes} we utilize an approximate posterior $q_\theta(z|x^{t:t+\delta})$ during training. We optimize our model using a modified $\beta$-VAE loss \cite{Higgins2016Beta-vae:Framework,Ilse2020DIVA:Autoencoders}. In our experiments with the auxiliary regressor, Mean Squared Error (MSE) is used to optimize the regressor. The contribution of the MSE to the overall loss is scaled using a hyperparameter $\beta_{\text{aux}}$. The complete loss term for VADA is given by Equation \ref{eq:loss} and the loss in the experiments with the auxiliary regressor is given by Equation \ref{eq:aux_loss}. 

\begin{align}
\label{eq:loss}\mathcal{L}_{\mathrm{VADA}}(x^{t:t+\delta}, x^{t-\delta:t}, y^{t:t+\delta}) ={}& -\mathbb{E}_{q_\theta(z\mid x^{t:t+\delta})}\big[\log p_\theta(x^{t:t+\delta}\mid z, x^{t-\delta :t}) \big]\\
&+\beta_{KL}\,D_{KL}\left[q_\theta(z\mid x^{t:t+\delta}) \,{\mid\mid}\, p_\theta(z\mid y')\right],\nonumber\\
\label{eq:aux_loss} \mathcal{L}_{aux}(x^{t:t+\delta}, x^{t-\delta:t}, y^{t:t+\delta}) ={}& \mathcal{L}_{\mathrm{VADA}}(x^{t:t+\delta}, x^{t-\delta:t}, y^{t:t+\delta}) \\
&+ \mathbb{E}_{q_\theta(z|x^{t:t+\delta})}\big[\beta_{aux} (y'-\hat{y}')^2 \big]\nonumber
\end{align}
where $y'={}\sum_t f_\theta(y^{t:t+\delta})/\delta$ and $\hat{y}'={}r_\theta(z)$.

\section{Experiments}
The goals of this empirical analysis are to \textbf{1)} asses VADA's effectiveness in modeling the distribution over nanopore currents, \textbf{2)} compare our data-driven approach to existing methods for nanopore simulation, \textbf{3)} determine whether the learned latent space can be used for the analysis of underlying sources of variation in nanopore current measurements.

A dataset provided by Oxford Nanopore Technologies is used for training VADA and performing the experiments. The dataset consists of sequenced human DNA and is publicly available for download on \href{https://github.com/nanoporetech/bonito}{\UrlFont{GitHub}}\footnote{\url{https://github.com/nanoporetech/bonito}}. The nanopore current sequences are split into sequences of length $1000$, resulting in a total of $1089009$ sequences. A test set containing $5\%$ of the available sequences is used for assessing the performance of VADA and a separate training set is used for training all models. The network architectures can be found in Appendix \ref{appendix} and code is available on \href{https://github.com/jmniederle/VADA}{\UrlFont{GitHub}}\footnote{\url{https://github.com/jmniederle/VADA}}.

\subsection{Simulation Evaluation}
\begin{figure}
\centering
\includegraphics[width=\textwidth]{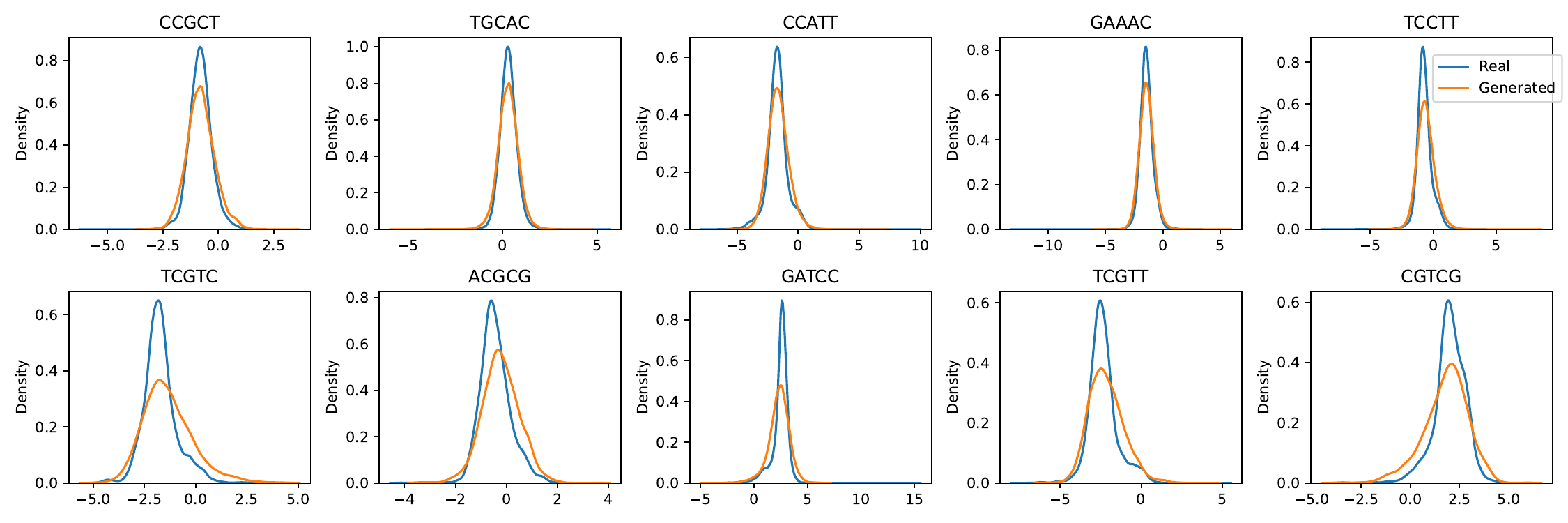}
\caption{Qualitative results for well-performing k-mers (top) and under-performing k-mers (bottom).} \label{fig:kmer_dist}
\end{figure}

We simulate nanopore currents for all sequences in our test set and visualize the true and simulated current distribution for different k-mers. Specifically, Figure \ref{fig:kmer_dist} shows simulation results for several k-mers where VADA produces current distributions that accurately match the distribution in experimental data (top row Fig. \ref{fig:kmer_dist}) and samples with worse simulation results (bottom row Fig. \ref{fig:kmer_dist}). From these qualitative results, we conclude that indeed the variability and distribution of nanopore current measurements differ between k-mers. Note that several k-mers on the bottom row contain the bases \textit{CG}, a combination of bases for which methylation commonly occurs \cite{Pages-Gallego2023ComprehensiveBasecalling}. Methylation influences the resulting current measurements and can potentially explain the skewed distribution shape. On the other hand, VADA produces a distribution that closely matches the distribution of each k-mer. However, the results are not yet perfect, and for most k-mers, VADA underestimates the mode of the distribution, and for some k-mers, the tails of the distribution are not precisely modeled. 

Besides investigating the results qualitatively, we quantitatively measure the performance of VADA's simulations and compare the results to a state-of-the-art nanopore simulation approach that is not fully data-driven. We wish to compare simulated distributions to experimental nanopore current data, to quantify VADA's effectiveness in capturing k-mer-specific variability. We use the Kolmogorov-Smirnov (KS) test statistic $D_{KS}$ as a metric to compute the similarity between the distribution of measurement values conditioned on the k-mer for simulated samples versus the corresponding distribution of experimental samples. 

$D_{KS}$ measures the largest absolute difference between the cumulative distribution functions, as such $D_{KS}=0$ means the distributions correspond perfectly and $D_{KS}=1$ means the distributions do not overlap at all. It makes no assumptions on either distribution being tested, allows for a two-sample test, and is widely used accross many disciplines.

We measure $D_{KS}$ for each individual k-mer using VADA and the existing non-data-driven DeepSimulator \cite{Li2020DeepSimulator1.5:Sequencing} and report the results in the first column of Table \ref{tab:sim_res}. VADA performs competitively to DeepSimulator in terms of $D_{KS}$, thus simulations sampled from VADA match the distribution of nanopore currents of experimental data as well as simulations sampled from DeepSimulator. Furthermore, the standard deviation in terms of $D_{KS}$ is much lower for VADA compared to DeepSimulator. From this, we conclude that VADA's ability to produce an accurate distribution is more stable across k-mers compared to DeepSimulator.


\begin{table}[b]
\caption{Evaluation of VADA and DeepSimulator in terms of simulation performance and base classification performance. We report mean $\pm$ standard deviation of $D_{KS}$ (lower is better) over all k-mers and AUC (higher is better). For each metric, we report score (standard deviation) over five independent runs.}\label{tab:sim_res}
\centering
\begin{tabular}{l|l|l}
Model         & Simulation Performance ($D_{KS}$) & Base Classification (AUC) \\ \hline
DeepSimulator \cite{Li2020DeepSimulator1.5:Sequencing} & $\mathbf{0.10}\ (0.000) \pm 0.06\ (0.000)$ & -\\
VADA (unconditional prior) & $0.32\ (0.001) \pm 0.10\ (0.003)$&$0.82\ (0.002) $\\
VADA (ours) & $\mathbf{0.13}\ (0.003) \pm 0.03\ (0.002)$ & $\mathbf{0.92}\ (0.000)$\\
VADA (with regressor)   & $\mathbf{0.13}\ (0.002) \pm 0.03\ (0.001)$ & $\mathbf{0.92}\ (0.001)$
\end{tabular}

\end{table}





\subsection{Analysis of Latent Space}
\label{sec:analysis_of_latent_space}
Furthermore, to investigate if the learned latent space can be used for the analysis of underlying sources of variation, we investigate if we can perform multi-label classification on the latent space. Specifically, we first sample a latent sample $z$ by using the approximate posterior, i.e. the encoder, where intuitively $z$ might contain a description of the sources of variation that resulted in nanopore current measurements $x^{t:t+\delta}$. Together with the embedding matrix $\theta_f$ of the embedding layer, $z$ is processed by a classifier $g(z, \theta_f)$ to produce a binary class prediction for every possible k-mer. 

There are $1025=4^5+1$ classes in total, representing all possible k-mer combinations of length $5$ and a class for incomplete k-mers, which can occur at the start or end of a nanopore current sequence. We measure performance by computing the Area Under the ROC (AUC) on the test set and report a resulting AUC of $\mathbf{0.92}$.

We conclude that VADA has learned an informative latent representation that can be used for determining the k-mers that correspond to a window of nanopore measurements, thereby, successfully extracting one of the underlying sources of variation. 


\subsubsection{Importance of the conditional prior} 
To show that the conditional prior is indeed the right tool for creating an informative latent space, we perform an ablation analysis where we compare to a model that does not use a conditional prior, and to one that, on top of using a conditional prior, is trained together with an auxiliary regressor to potentially make the latent-space even more informative for basecalling. 

In the model with the unconditional prior, we use an unconditional gaussian $\mathcal{N}(0, 1)$ and condition the decoder $p_\theta(x^{t:t+\delta}|z, x^{t-\delta:t}, y')$ and approximate posterior $q_\theta(z|x^{t:t+\delta}, y')$ on the aggregated k-mer embedding $y'$, instead. 

To train the model with the auxiliary regressor, we use the loss shown in \eqref{eq:aux_loss}.  

Details on the training procedure can be found in Appendix \ref{appendix}. From the results in Table \ref{tab:sim_res}, we conclude that indeed conditioning via the prior is essential for achieving simulation performance competitive to the non-data-driven DeepSimulator. Neglecting the prior in the conditioning mechanism clearly results in worse simulation performance and in a much less interpretable latent-space. On the otherhand, using an auxiliary regressor to further encourage an informative latent-space seems to provide no added value over using just the conditional prior.

\section{Conclusions}
In this work, we proposed VADA, a fully data-driven approach for probabilistic simulation of nanopores. We evaluate VADA on experimental nanopore sequencing data and demonstrate that VADA performs competitively with existing non-data-driven approaches. Furthermore, we show VADA has learned a meaningful latent representation by demonstrating that we can accurately classify k-mers from this representation. To conclude, we demonstrate the value of deep generative models for simulating nanopore sequencing and capturing the underlying sources of variability for nanopore currents.

\subsection{Future Research}
Although VADA does not outperform non-data-driven approaches in simulation quality outright, we do manage to match their performance without exploring the many techniques that have been developed over the years for improving the performance of VAE-like models \cite{Tschannen2018RecentLearning}. Extending VADA with such techniques could be a fruitful direction of further research into nanopore sequencing simulation.

As we have demonstrated, the learned latent representation can capture useful information about the DNA being sequenced in experimental data. Further investigations into what sources of variability are captured could lead to new analytic methods for nanopore signals. Such investigations may enable easier detection of chemical modifications of DNA bases, leading to better understanding of their role in DNA function.

Various methods have been developed for separating sources of variability in latent space models \cite{Ilse2020DIVA:Autoencoders,Tschannen2018RecentLearning}. Such techniques might improve the simulation capability of VADA, and enable the use of the encoder for more downstream tasks through added structure in the latent space.

\bibliographystyle{splncs04}
\bibliography{references}






\appendix

\section{VADA Model Architecture}\label{appendix}

An embedding size of $64$ is used for the embedding layer $f_\theta$, the latent samples have size $32$ and the prediction window size is set to $16$. Models are trained for $140000$ training steps, using Adam \cite{Kingma2014Adam:Optimization} with batch size $512$. Loss hyperparameters are set to $\beta_{aux}=0.03$ and $\beta_{KL}=0.005$.

We use a ResBlock(in\_channels, out\_channels), consisting of: 

\noindent Conv1D(out\_channels, 3), BatchNorm1D, Conv1D(out\_channels, 3), BatchNorm1D, and finally a residual (skip) connection \cite{He2016DeepRecognition} of the input to the result. We let the parameters of Conv1D represent (output channels, kernel size). Furthermore, ConvTranspose1D and Conv1D are used to, respectively, upsample and downsample the input.
\begin{table}
\centering
\caption{Architectures for the neural networks used in our experiments.}
\begin{subtable}[t]{.45\textwidth}
\caption{Architecture for the prior $p_\theta(z|y')$. The model has two heads, one for the mean and one for the scale. 
}\label{tab:arch_prior}
    \centering
    \begin{tabular}{l|l}
    Block   & Details \\ \hline
      1     & Linear(64, 64), LeakyReLU  \\
      2   & Linear(64, 64), LeakyReLU \\
        3-$\mu$  & Linear(64, 32) \\
      3-$\sigma$   & Linear(64, 32), Softplus \\
    \end{tabular}
\end{subtable}
\hspace{.05\textwidth}
\begin{subtable}[t]{.45\textwidth}
\caption{Architecture for the auxiliary regressor $r_\theta(z)$.}\label{tab:arch_aux}
    \centering
    \begin{tabular}{l|l}
    Block   & Details \\ \hline
      1     & Linear(32, 64), LeakyReLU  \\
      2   & Linear(64, 64), LeakyReLU \\
    3  & Linear(64, 32) \\
    \end{tabular}
\end{subtable}
\vspace{10pt}
\begin{subtable}[t]{.45\textwidth}
\caption{Architecture for decoder $p_\theta(x^{t:t+\delta}|z, x^{t-\delta:t})$. The network outputs values for the mean and scale is fixed at $1/\sqrt{2}$.
}\label{tab:arch_decoder}
    \centering

     \begin{tabular}{l|l}
    Block   & Details \\ \hline
      1     & Linear(48, 64), LeakyReLU  \\
      2   & ResBlock(64, 64) \\
      3 & ResBlock(64, 32) \\
      4 & UpSample \\
      5   & ResBlock(32, 32)\\
      6 & ResBlock(32, 16)\\
      7 & UpSample \\
      8 & ResBlock(16, 16)\\
      9 & ResBlock(16, 8) \\
      10 & UpSample\\
        11 & ResBlock(8, 8) \\
        12 & ResBlock(8, 4) \\
        13 & UpSample\\
      14 & Conv1D(1, 1)
    \end{tabular}
    

\end{subtable}
\hspace{.05\textwidth}
\begin{subtable}[t]{.45\textwidth}

\caption{Architecture for encoder $q_\theta(z|x^{t:t+\delta})$. The distribution is Gaussian, with the following parameterisation.
}\label{tab:arch_encoder}
    \centering

    \begin{tabular}{l|l}
    Block   & Details \\ \hline
      1     & Conv1d(4, 1)  \\
      2   & ResBlock(4, 8)\\
      3 & ResBlock(8, 8) \\
      4 & DownSample \\
      5   & ResBlock(8, 16) \\
      6 & ResBlock(16, 16) \\
      7 & DownSample \\
      8 & ResBlock(16, 32) \\
      9 & ResBlock(32, 32) \\ 
      10 & DownSample\\
      11 & ResBlock(32, 64) \\
      12 & ResBlock(64, 64) \\ 
      13 & DownSample\\
      14-$\mu$ & Linear(64, 64), LeakyReLU \\ 
      14-$\sigma$ & Linear(64, 64), LeakyReLU\\
      15-$\mu$ & Linear(64, 32) \\
      15-$\sigma$ & Linear(64, 32), Softplus
    \end{tabular}

\end{subtable}

\end{table}

\end{document}